\documentclass[11pt, a4paper]{article}
\usepackage{jcappub, bm}

\newcommand{\nhat}{\mathbf{\hat{n}}}
\newcommand{\vbulk}{\mathbf{v}_\text{bulk}}
\newcommand{\vb}{v_\text{bulk}}
\newcommand{\chit}{\tilde{\chi}}

\title{No evidence for bulk velocity from type Ia supernovae}

\author[a,b,c]{Dragan Huterer,}
\author[a]{Daniel L. Shafer}
\author[b]{and Fabian Schmidt}

\affiliation[a]{Department of Physics, University of Michigan,\\
450 Church Street, Ann Arbor, MI 48109, U.S.A.}
\affiliation[b]{Max-Planck-Institut f\"ur Astrophysik,\\
Karl-Schwarzschild-Str.\ 1, 85748 Garching, Germany}
\affiliation[c]{Excellence Cluster Universe,\\
Boltzmannstr.\ 2, D-85748 Garching, Germany}

\emailAdd{huterer@umich.edu}
\emailAdd{dlshafer@umich.edu}
\emailAdd{fabians@mpa-garching.mpg.de}

\abstract{We revisit the effect of peculiar velocities on low-redshift type Ia
  supernovae. Velocities introduce an additional guaranteed source of
  correlations between supernova magnitudes that should be considered in all
  analyses of nearby supernova samples but has largely been neglected in the
  past. Applying a likelihood analysis to the latest compilation of nearby
  supernovae, we find no evidence for the presence of these correlations,
  although, given the significant noise, the data is also consistent with the
  correlations predicted for the standard $\Lambda$CDM model. We then consider
  the dipolar component of the velocity correlations --- the frequently studied
  ``bulk velocity'' --- and explicitly demonstrate that including the velocity
  correlations in the data covariance matrix is crucial for drawing correct
  and unambiguous conclusions about the bulk flow. In particular, current
  supernova data is consistent with no excess bulk flow on top of what
  is expected in $\Lambda$CDM and effectively captured by the covariance. We
  further clarify the nature of the apparent bulk flow that is inferred when
  the velocity covariance is ignored. We show that a significant fraction of
  this quantity is expected to be noise bias due to uncertainties in supernova
  magnitudes and not any physical peculiar motion.}


\begin{document}
\maketitle

\section{Introduction}
\label{sec:intro}

Motions of objects in the universe are not entirely random. Objects which are
physically close to one another respond similarly to the pull of large-scale
structure, and as a result their peculiar velocities are
correlated. Correlations between galaxy peculiar velocities are an old subject
\cite{Kaiser:1989kb,Gorski_etal,Sandage:1979re,Nusser:2011tu,Macaulay:2011av,Branchini:2012rb,Feix:2014bma},
and these velocities have already been used to constrain cosmological models,
in particular the amount of matter in the universe (see \cite{Strauss:1995fz}
for a review). More recently, peculiar velocities have become important in the
analysis of type Ia supernova (SN Ia) data. At low redshift ($z \lesssim
0.05$), typical peculiar velocities of $\sim$300~km/s are a significant
contribution to the SN redshift (for instance, $cz =$ 3,000~km/s at $z =
0.01$). These peculiar velocities are a nuisance if one is interested in using
the SNe to constrain expansion history and dark energy, and it is common
practice to propagate this extra dispersion into the error budget (e.g.\ add
$300~\text{km/s} \times 5/(cz \ln10)$ in quadrature to the statistical
uncertainty of each SN magnitude). However, this neglects significant
covariance between the velocities of different SNe.

Alternatively, one can consider the SN peculiar velocity field itself to be a
signal, one that should contain useful information about the amount and
distribution of matter in the universe. Nearby SNe are much fewer in number
than nearby galaxies, and given the volume limitation for both, this will
likely still be the case in the future. On the other hand, SNe are more useful
on a per-object basis because their individual distances can be inferred
directly and with relative precision --- roughly $7\%$ for each SN, depending
on the quality of the observations. Therefore, there has been a resurgence of
interest in how SN peculiar velocities are modeled and used \cite{Hui:2005nm, Davis:2010jq}.

In this paper, we perform a careful study of the SN velocity correlations in
current data, in particular the way in which they are used to draw conclusions
about the so-called ``bulk velocity'' --- the motion, relative to the cosmic
microwave background (CMB) rest frame, of the patch of the universe centered
on us and containing the nearby sample. Though we focus on SNe, our
methodology is not restricted to SNe and equally applies to analysis of galaxy
peculiar velocities.

The paper is organized as follows. In section~\ref{sec:theory}, we review the
physics of how peculiar velocities affect SN magnitudes. In
section~\ref{sec:data}, we describe the SN samples and how we use them. In
section~\ref{sec:like}, we define a general likelihood that is the basis for
our analyses, which include a test for the presence of velocity correlations
(section~\ref{sec:A}), a test for the presence of excess bulk velocity beyond
that encoded in the correlations predicted in $\Lambda$CDM
(section~\ref{sec:vbulk}), and a comparison to previous work that studied bulk
flows without the velocity covariance (section~\ref{sec:vbulk_noise}). We
summarize our conclusions in section~\ref{sec:concl}.

\section{Theoretical framework}
\label{sec:theory}

\subsection{Magnification and SN magnitude residuals at low redshifts}

The magnitude residuals of standard candles like SNe Ia are directly related
to the magnification $\mu$, which is defined as the fractional perturbation in
the angular diameter and luminosity distances (see \cite{Hui:2005nm,
  Bonvin:2005ps}),
\begin{equation}
-\frac{1}{2} \, \mu = \frac{\Delta d_L}{\bar{d}_L(z)} = \frac{\Delta d_A}{\bar{d}_A(z)} \, ,
\label{eq:Mdef}
\end{equation}
where $\bar{d}_A(z)$ and $\bar{d}_L(z)$ denote the background distances
evaluated at the observed redshift $z$. The second equality, relating
luminosity and angular diameter distances, follows from the conservation of
the photon phase space density. That is, $\mu$ describes both the change in
the apparent angular size of a spatial ruler as well as the change in observed
flux of a standard candle.

Covariant expressions for the magnification at linear order in cosmological
perturbations have been given in \cite{Bonvin:2005ps, Challinor:2011bk,
  Bonvin:2011bg, Schmidt:2012ne}. In the conformal-Newtonian (cN) gauge, where
the metric is written as
\begin{equation}
ds^2 = a^2(\tau) \left[- (1 + 2\Psi) \, d\tau^2 + (1 + 2\Phi) \, \delta_{ij} \, dx^i dx^j \right] \, ,
\label{eq:metric_cN}
\end{equation}
the magnification is given, in the notation of \cite{Schmidt:2012ne},
by\footnote{Here, we have neglected a term that is present if the luminosity
  of the standard candle depends on time; in any case, it is subdominant in
  the limit we will consider. We have also neglected two pure monopole
  contributions, motivated by the discussion in section~\ref{sec:monopole}.}
\begin{equation}
\mu = \left[- 2 + \frac{2}{a H \chit} \right] \Delta\ln a - 2\Phi + 2\kappa - 2v_{\parallel o} - \frac{2}{\chit} \int_0^{\chit} d\chi \: (\Psi - \Phi) \, ,
\label{eq:magcN}
\end{equation}
where $\chit \equiv \chi(z)$ is the coordinate distance inferred using the
observed redshift, and
\begin{align}
\kappa &= \frac{1}{2} \int_0^{\chit} d\chi \, \frac{\chi}{\chit} (\chit - \chi) \nabla_\perp^2 \left(\Psi - \Phi \right) \, , \label{eq:kcN} \\
\Delta\ln a &= \Psi_o - \Psi + v_\parallel - v_{\parallel o}
+ \int_0^{\chit} d\chi \left[\Phi' - \Psi' \right] \label{eq:Dlna_cN}
\end{align}
are the convergence and fractional redshift perturbation, respectively. The
latter contains the gravitational redshift, Doppler shift, and integrated
Sachs-Wolfe effect. Further, $\nabla_\perp^2$ denotes the Laplacian on a
sphere of radius $\chi$, $v_\parallel = v^i \hat n_i$ denotes the peculiar
velocity projected along the line of sight $\nhat$, integrals over $\chi$
denote integrals along the past lightcone, and a subscript $o$ denotes a
quantity evaluated at the location of the observer. Note that $\kappa$,
$\Delta\ln a$, and any other terms appearing in eq.~\eqref{eq:magcN} are
coordinate-dependent quantities, and only the combination given in
eq.~\eqref{eq:magcN} corresponds to an actual (gauge-invariant)
observable. This can be verified by considering gauge transformations and
various test cases \cite{Jeong:2011as, Schmidt:2012ne}.

For low-redshift SNe, where $z \ll 1$ so that $\chit \ll 1/(a H)$, the terms
involving the velocity are the most significant. This is because the lensing
convergence is suppressed for small source distances and because the scales
probed are much smaller than the horizon. Then, the terms involving the
potentials $\Phi$ and $\Psi$ are also suppressed by roughly $a H \chit$
relative to the velocity. In this case, we obtain
\begin{equation}
\mu \stackrel{z \ll 1}{=} \frac{2}{a H \chit} \left(v_\parallel - v_{\parallel o} \right) \, .
\label{eq:Mlowz}
\end{equation}
This expression for the magnification, proportional to the relative velocity
along the line of sight between source and observer, simply arises due to the
fact that we evaluate the luminosity distance to a supernova using the
background distance-redshift relation, while the actual redshift is perturbed
by the Doppler shift. One can easily verify numerically that this
approximation is better than 1\% for $z \lesssim 0.1$, which is the redshift
range we will consider in this paper. Note that \cite{Hui:2005nm} includes the
term $-2$ in the $\Delta\ln a$ prefactor in eq.~\eqref{eq:magcN}; however,
this is not strictly consistent, since the terms involving $\Psi$ in
$\Delta\ln a$, as well as the aberration term $-2 v_{\parallel o}$ in
eq.~\eqref{eq:magcN} are of comparable magnitude to this correction (see also
\cite{Kaiser:2014jca}). We will thus work with eq.~\eqref{eq:Mlowz} as the
proper low-z limit of eq.~\eqref{eq:magcN}. Note that this relation remains
valid even if $\mu$ becomes of order unity, as long as the velocities
$v_\parallel$ remain small compared to the speed of light. On very small
scales ($z < 0.01$), the velocities are no longer described accurately by
linear perturbation theory.  However, since the SN samples considered here are
restricted to $z \gtrsim 0.01$, we will work with velocities derived from
linear perturbation theory. Note that, in principle, nonlinear
  corrections to the velocity could also be relevant for higher-redshift SNe,
  if two SNe happen to be physically close. However, we have verified that
  nonlinear corrections to the velocities have a small effect (see below).

As eq.~\eqref{eq:Mlowz} shows, the relevant quantity for the magnification at
low $z$ is the relative velocity between the source and the observer projected
along the line of sight. This also includes small-scale motions such as the
velocity of the Solar System with respect to the Milky Way center, which are
uncorrelated with large-scale cosmological velocity fields. For this reason,
we correct the observed SN redshifts to the CMB rest frame using the measured
CMB dipole moment (see section~\ref{sec:data}). Then, the magnification
becomes
\begin{equation} 
\mu|_{z_\text{CMB}} = \frac{2}{a H \chit} \left(v_\parallel - v_{\parallel, \text{CMB}} \right) \, ,
\end{equation}
where the relevant quantity is now the velocity of the SN relative to the CMB
rest frame. This simplifies the interpretation, since $v_\parallel -
v_{\parallel, \text{CMB}}$ is well described by linear perturbation theory. In
fact, by performing the calculation in the CMB rest frame (as is normally
done), we can set $\mathbf{v}_\text{CMB} = 0$. The following relations will
always assume that we work with CMB-frame redshifts and in the CMB rest frame.

It is straightforward to convert a perturbation in the luminosity distance (as
in eq.~\eqref{eq:Mdef}) into a perturbation of the SN magnitude from the
homogeneous background value:
\begin{equation}
  \delta m = - \frac{5}{2\ln 10} \,
  \mu \stackrel{z \ll 1}{=} - \frac{5}{\ln 10} \, \frac{\mathbf{v} \cdot \nhat}{a H \chit} \, .
\label{eq:deltam}
\end{equation}
Note that this relation assumes that $\mu \ll 1$ and is thus not applicable at
very low redshifts. While it is straightforward to derive the proper nonlinear
relation for $\delta m$, this is not necessary for our purposes, since $z
\gtrsim 0.01$ in our SN samples.

Now consider one object at redshift $z_i$ in direction $\nhat_i$ on the sky, and a
second at $(z_j, \nhat_j)$. We can derive the covariance of their residuals,
\begin{align}
  S_{ij} \equiv \left\langle \delta m_i \, \delta m_j \right\rangle &=
  \left[\frac{5}{\ln 10}\right]^2 \frac{a_i}{a'_i \chi_i} \,
  \frac{a_j}{a'_j \chi_j} \, \xi_{ij} \nonumber \\
  &= \left[\frac{5}{\ln 10}\right]^2 \frac{(1 + z_i)^2}{H(z_i) d_L(z_i)} \,
  \frac{(1 + z_j)^2}{H(z_j) d_L(z_j)} \, \xi_{ij} \, , \label{eq:Sij}
\end{align}
where $\xi_{ij}$ is the velocity covariance given by
\begin{align}
  \xi_{ij} \equiv \xi_{ij}^\text{vel} &\equiv
  \langle(\mathbf{v}_i \cdot \nhat_i) (\mathbf{v}_j \cdot \nhat_j) \rangle
  \nonumber \\[0.2cm]  
  &= \frac{dD_i}{d\tau} \, \frac{dD_j}{d\tau}
  \int\frac{dk}{2\pi^2} P(k, a = 1) \sum_\ell (2\ell + 1)
  j'_\ell(k\chi_i) j'_\ell(k\chi_j) P_\ell(\nhat_i \cdot \nhat_j) \, . \label{eq:xi}
\end{align}
Here, primes denote derivatives of the Bessel functions with respect to their
arguments, $\tau$ is the conformal time, $d\tau = dt/(a^2 H)$, $D_i$ is the
linear growth function evaluated at redshift $z_i$, and $\chi_i =
\chi(z_i)$. The power spectrum $P(k, a)$ is evaluated in the present ($a = 1$)
and, at the large scales we are interested in, only the first $\simeq 10$
terms in the sum over the multipoles contribute. As mentioned above, we
  use velocities derived from linear theory and thus insert the linear matter
  power spectrum for our numerical results. We have verified that using a
  prescription for the nonlinear matter power spectrum in eq.~\eqref{eq:xi}
  does not significantly affect our results. We thus conclude that the linear
  treatment is sufficient for our purposes. Physically, this is because the
  dominant contribution to the covariance comes from fairly large-scale
  modes. Note that in our approach, $\langle (\delta m_i)^2 \rangle$ is
assumed to capture the random motion contribution to the variance of SN
residuals.\footnote{An alternative approach by \cite{Neill:2007fh} models
  velocities with perturbation theory based on a density field derived from
  other surveys, and complements them with a ``thermal'' component of
  $\sim$150~km/s added in quadrature to account for nonlinearities. In
  contrast to our approach, this thus relies on external data sets.  A
  detailed comparison between the covariances obtained using these different
  approaches would be interesting but is beyond the scope of this paper.}
While this is not expected to be completely accurate when using the linear
matter power spectrum, the difference in the diagonal covariance elements is
not very significant.\footnote{For the low-redshift SNe we consider, nearly
  all of the redshifts are derived from host galaxy spectra, and so the motion
  of the SN within its host does not contribute to the residuals.}

We have denoted this covariance matrix $\mathbf{S}$ to emphasize that this is
a cosmologically guaranteed ``signal'' to be added to the ``noise'' covariance
matrix that accounts for the combination of statistical and systematic errors
that affects SN distance measurements, such as intrinsic variations in the SN
luminosity (see section~\ref{sec:data}). We again point out that the two
geometric prefactors in eq.~\eqref{eq:Sij} each differ by an additive factor
of $1$ relative to those in \cite{Hui:2005nm} because we drop the term $-2$ in
eq.~\eqref{eq:magcN} in order to achieve a consistent low-redshift expansion;
we have checked that all neglected terms would contribute negligibly at $z
\lesssim 0.1$.

\subsection{Monopole subtraction}
\label{sec:monopole}

The magnification eq.~\eqref{eq:Mdef} and its low-redshift version
eq.~\eqref{eq:Mlowz} still have a monopole component, that is, a contribution
that is uniform on the sky. However, since the SN magnitude residuals are
defined with respect to the best-fit distance-redshift relation, this monopole
is mostly absorbed in the fit. While there could technically be a residual
monopole signal due to the fact that our fit (to a flat $\Lambda$CDM model,
see section~\ref{sec:like}) is very restricted, we will assume here that the
bulk of the monopole is removed. Thus, eq.~\eqref{eq:Sij} needs to be
corrected.

To this end, we define the mean magnitude residual at redshift $i$ as
\begin{equation}
\overline{\delta m}(z_i) = \int \delta m(z_i, \nhat) \, W(\nhat) \, d^2\nhat \, ,
\end{equation}
where $W(\nhat)$ is the survey window function, which is normalized such that
$\int W(\nhat) \, d^2\nhat = 1$. Then, noting that we actually measure
$\widehat{\delta m}_i = \delta m_i - \overline{\delta m}(z_i)$, the proper
covariance is
\begin{equation}
S_{ij} \equiv \left\langle \widehat{\delta m}_i \, \widehat{\delta m}_j \right\rangle = \left\langle \left[\delta m(z_i, \nhat_i) - \overline{\delta m}(z_i) \right] \left[\delta m(z_j, \nhat_j) - \overline{\delta m}(z_j) \right] \right\rangle \, ,
\end{equation}
which can be worked out to be
\begin{align}
S_{ij} &= \left[\frac{5}{\ln 10} \right]^2 \frac{(1 + z_i)^2}{H(z_i) d_L(z_i)} \, \frac{(1 + z_j)^2}{H(z_j) d_L(z_j)} \, \frac{dD_i}{d\tau} \, \frac{dD_j}{d\tau} \int \frac{dk}{2\pi^2} \, P(k, a = 1) \nonumber \\
&\times \sum_\ell (2\ell + 1) j'_\ell(k \chi_i) j'_\ell(k \chi_j) \left[P_\ell(\nhat_i \cdot \nhat_j) - \frac{4\pi}{2\ell + 1} \left[w_\ell(\nhat_i) + w_\ell(\nhat_j) \right] + 4\pi W_\ell \right],
\label{eq:Sij_with_window}
\end{align}
where the survey footprint has been expanded in spherical harmonics,
\begin{equation}
W(\nhat) = \sum_{\ell m} w_{\ell m} Y_{\ell m}(\nhat) \, ,
\end{equation}
and the coefficients $w_\ell(\nhat_i)$ and $W_\ell$ are defined as
\begin{equation}
w_\ell(\nhat) \equiv \sum_m w_{\ell m} Y_{\ell m}(\nhat) \, , \qquad W_\ell \equiv \frac{\sum_m |w_{\ell m}|^2}{2\ell + 1} \, .
\end{equation}
The extra terms in the square brackets in eq.~\eqref{eq:Sij_with_window} are
corrections due to the survey window. The $W_\ell$ are therefore just the
angular power spectrum (more precisely, the ``pseudo-$C_\ell$'') of the map,
while $w_\ell(\nhat)$ is the $\ell$ portion of the survey mask at an arbitrary
location. Note that, due to the required normalization of $W$, its value where
the survey observes is not unity, but rather
\begin{equation}
W(\nhat) = \left\{
\begin{array}{cc}
\displaystyle \frac{1}{\Omega_\text{sky}} & \text{(observed sky)} \\[0.4cm]
0 & \text{(unobserved sky)}.
\end{array}
\right.
\label{eq:Wnhat}
\end{equation}

The term in the square parentheses in the last line of
eq.~\eqref{eq:Sij_with_window}, which includes the subtraction of the mean, is
therefore a new result that has not, to our knowledge, been derived and
included in previous analyses (although the existence of such a term has been
pointed out in \cite{Hui:2005nm, Kaiser:2014jca}). For a full-sky window, it
is easy to show that this term becomes $P_\ell(\nhat_i \cdot \nhat_j) - 1$ for
$\ell = 0$ and remains equal to the original expression $P_\ell(\nhat_i \cdot
\nhat_j)$ for the other multipoles.

We find that the monopole-subtracted formula leads to small but noticeable
changes in the results, such as the constraints on the parameter $A$ in
section~\ref{sec:A}, and we recommend that it be used in future analyses.

\section{SN Ia data and noise covariance}
\label{sec:data}

For our primary SN Ia dataset, we use the joint light-curve analysis (JLA)
\cite{Betoule:2014frx} of SNe from the Supernova Legacy Survey (SNLS) and the
Sloan Digital Sky Survey (SDSS). JLA includes a recalibration of SNe from the
first three years of SNLS \cite{Conley:2011ku, Betoule:2012an} along with the
complete SN sample from SDSS, making it the largest combined SN analysis to
date. The final compilation includes 740~SNe, $\sim$100 low-redshift SNe from
several subsamples, $\sim$350 from SDSS at low to intermediate redshifts,
$\sim$250 from SNLS at intermediate to high redshifts, and $\sim$10
high-redshift SNe observed with the Hubble Space Telescope.

We combine the individual covariance matrix terms
provided\footnote{\url{http://supernovae.in2p3.fr/sdss_snls_jla/}} to compute
the full covariance matrix, which includes statistical errors (correlated
uncertainties in the light-curve measurements, intrinsic scatter, lensing
dispersion) and a variety of systematic errors (photometric calibration,
uncertainty in the bias correction, light-curve model uncertainty, potential
non-Ia contamination, uncertainty in the Milky Way dust extinction correction,
and uncertainty in the host galaxy correction).

Although we compute the covariance matrix as described in the JLA analysis, we
leave out two contributions to the total error. First, we leave out the
additional scatter of $150~\text{km/s} \times 5/(cz \ln 10)$ added in
quadrature to the other statistical errors on the diagonal to account for
peculiar velocity. This peculiar velocity scatter does not apply because
peculiar velocities are not a source of noise in our analysis; instead, they
are modeled by the formalism discussed in section~\ref{sec:theory}. We also
leave out the systematic error term corresponding to uncertainty in the
peculiar velocity \emph{correction} applied to the low-$z$ JLA
redshifts. Since our aim is to study the peculiar velocities themselves, we
want to avoid this correction and then leave out the systematic error
associated with it. To this end, we obtain the CMB-frame redshifts
$z_\text{CMB}$ directly from the measured heliocentric redshifts
$z_\text{hel}$. Specifically, for each SN we compute
\begin{equation}
1 + z_\text{CMB} = (1 + z_\text{hel}) \left[1 + \frac{v_\text{CMB}}{c} \, (\nhat_\text{CMB} \cdot \nhat) \right] \, ,
\label{eq:zhel2zcmb}
\end{equation}
where $\nhat$ is the sky position of the SN, $\nhat_\text{CMB}$ is the CMB
dipole direction, and $v_\text{CMB}$ is the velocity of the Solar System
barycenter relative to the CMB rest frame implied by the dipole amplitude. We
use the measured values $v_\text{CMB} = 369~\text{km/s}$ and $\nhat_\text{CMB}
\equiv (l, b) = (263.99^\circ, 48.26^\circ)$, where the quoted uncertainties
\cite{Hinshaw:2008kr} are negligible for our purposes.

For comparison, we separately consider the Union2 SN Ia analysis
\cite{Amanullah:2010vv} from the Supernova Cosmology
Project.\footnote{\url{http://supernova.lbl.gov/Union/}} We use the full covariance matrix provided for the
Union2 SN magnitudes, but as with JLA, we remove the peculiar velocity scatter
(300~km/s here) that was added to the diagonal. The redshifts given for the
Union2 SNe are just the heliocentric redshifts transformed to the CMB rest
frame, so we use them directly.

Note that the Union2 compilation of 557~SNe has been superseded by the
  Union2.1 compilation \cite{Suzuki:2011hu} of 580~SNe, but here the goal is a
  fair comparison to previous work that analyzes the Union2 data. Since the
  primary change in Union2.1 is the addition of a set of high-redshift SNe,
  and since only low-redshift SNe are relevant for our analysis, we would
  expect the two compilations to produce very similar results. When
  substituting Union2.1 for Union2, our results do not change qualitatively,
  but there are some minor differences due to new estimates for some corrected
  SN magnitudes and their errors. Union2.1 also includes a host-mass
  correction (see below) that Union2 does not, but this is relatively small
  and only accounts for part of the magnitude differences.

Finally, we briefly consider the low-$z$ compilation from the older analysis
of \cite{Jha:2006fm}, also for comparison with other work. This compilation
does not include an analysis of systematic errors, so the uncertainty in a SN
magnitude is just a combination of the light-curve measurement errors and the
derived intrinsic scatter of $\sigma_\text{int} = 0.08~\text{mag}$. Again, we
do not include the peculiar velocity scatter of 300~km/s prescribed for a
cosmological analysis. Although CMB-frame redshifts are given, we transform
the given heliocentric redshifts into CMB-frame redshifts ourselves using
eq.~\eqref{eq:zhel2zcmb}.

Because SNe Ia are not perfect standard candles, it is necessary to correct
the observed peak magnitude of each SN for the empirical correlations between
the SN Ia absolute magnitude and both the stretch (broadness) and color
measure associated with the light-curve fitter. More recently, it has become
common to fit for a constant offset in the absolute magnitude for SNe in
high-stellar-mass host galaxies. For JLA, the corrected magnitude is therefore given by
\begin{equation}
m^\text{corr} = m + \alpha \times (\text{stretch}) - \beta \times (\text{color}) + P \, \Delta M,
\end{equation}
where $\alpha$, $\beta$, and $\Delta M$ are nuisance parameters describing,
respectively, the stretch, color, and host-mass corrections. The measured $P
\equiv P(M_* > 10^{10} M_\odot)$ is the probability that the SN occurred in a
high-stellar-mass host galaxy. Note that, as mentioned above, Union2 does not include this host-mass correction; also, the analysis of \cite{Jha:2006fm} uses a different light-curve fitter with different (but related) light-curve corrections.

For each of the three datasets, we fix the SN Ia nuisance parameters to their
best-fit values from a fit to the Hubble diagram (for JLA, we perform this fit
and correct the magnitudes ourselves; for the other datasets, we use
precorrected magnitudes). In a proper cosmological analysis, one should vary
the SN nuisance parameters simultaneously with any cosmological parameters. In
practice, however, the nuisance parameters are well-constrained by the Hubble
diagram with little dependence on the cosmological model, so holding them
fixed should be a good approximation, especially for our purposes here.

\section{Likelihood}
\label{sec:like}

We write the full covariance $\mathbf{C}$ as the sum of two contributions,
$\mathbf{C} = \mathbf{S} + \mathbf{N}$, where $\mathbf{S}$ is the signal
covariance, dominated by velocities at low $z$ and discussed in
section~\ref{sec:theory}, and $\mathbf{N}$ is the noise covariance, described
in section~\ref{sec:data}.

Assuming a given cosmological model that allows us to calculate $\mathbf{S}$
(eq.~\eqref{eq:Sij}), the \emph{optimal} way to determine whether the data
favors peculiar velocities is to consider evidence for the detection of the
full signal matrix $\mathbf{S}$. This approach uses more information in the
data than the search for any particular moment, such as the dipole, of the
peculiar velocity field.

Here we would like to detect evidence for coherent departures of supernova
magnitudes from the mean --- that is, clustering. To do this, we introduce a
new dimensionless parameter $A$ and let $\mathbf{S} \rightarrow A \,
\mathbf{S}$, where $A = 1$ for the fiducial model. $A = 0$ corresponds to the
case that magnitude residuals are purely due to noise and systematics in the
SN Ia data. We would like to test whether $A$ is consistent with one and
different from zero. Including the new parameter $A$, the full covariance
becomes
\begin{equation}
\mathbf{C} = A \, \mathbf{S} + \mathbf{N} \ ,
\label{eq:Cov}
\end{equation}
where $0 \leq A < \infty$.

On the other hand, allowing for an excess bulk flow component is interesting as well,
as it can be used to search for signatures beyond the fiducial $\Lambda$CDM
model and also allows us to compare our results with the existing literature
(see section~\ref{sec:vbulk}). In this case, the magnitude residuals are
affected by an additional bulk velocity $\vbulk$ (e.g.\ \cite{Hui:2005nm}):
\begin{equation}
  \Delta m_i^\text{bulk} \equiv \Delta m^\text{bulk}(\vbulk; z_i, \nhat_i) =
  -\left(\frac{5}{\ln 10} \right) \, \frac{(1 + z_i)^2}{H(z_i) d_L(z_i)} \ \nhat_i \cdot \vbulk \, ,
\label{eq:dm}
\end{equation}
where $z_i$ and $\nhat_i$ are the redshift and sky position of the SN, while
$\vbulk$ is a fixed three-dimensional vector. In the quasi-Newtonian picture,
$\vbulk$ corresponds to the bulk motion of the SN sample; however, in the
context of non-standard cosmological models (such as those breaking
homogeneity or isotropy), this should really be seen as a convenient
parametrization of the dipole of SN magnitude residuals.

Putting these ingredients together, we construct a multivariate Gaussian
likelihood\footnote{Note that SN flux, or a quantity linearly related to it,
  might be a better choice for the observable than the magnitude, given that
  we expect the error distribution of the former to be more Gaussian than the
  latter. Nevertheless, this choice should not impact our results, as the
  fractional errors in flux are not too large, and we have explicitly checked
  that the distribution of the observed magnitudes around the mean is
  approximately Gaussian. Therefore we follow most literature on the subject
  and work directly with magnitudes.}
\begin{equation}
\mathcal{L}(A, \vbulk) \propto \frac{1}{\sqrt{|\mathbf{C}|}} \exp\left[-\frac{1}{2} \mathbf{\Delta m}^\intercal \mathbf{C}^{-1} \mathbf{\Delta m} \right] ,
\label{eq:like}
\end{equation}
where the elements of the vector $\mathbf{\Delta m}$ are
\begin{equation}
(\mathbf{\Delta m})_i = m^\text{corr}_i - m^\text{th}(z_i, \mathcal{M}, \Omega_m) - \Delta m^\text{bulk}_i(\vbulk) \, ,
\label{eq:dmvec}
\end{equation}
where $m^\text{corr}_i$ are the observed, corrected magnitudes and
$m^\text{th}(z_i, \mathcal{M}, \Omega_m)$ are the theoretical predictions for
the background cosmological model (see below). The $\mathcal{M}$ parameter
corresponds to the (unknown) absolute calibration of SNe Ia; we analytically
marginalize over it in all analyses (e.g.\ appendix of \cite{Shafer:2013pxa}).

\begin{figure}[t]
\centering
\includegraphics[width=\textwidth]{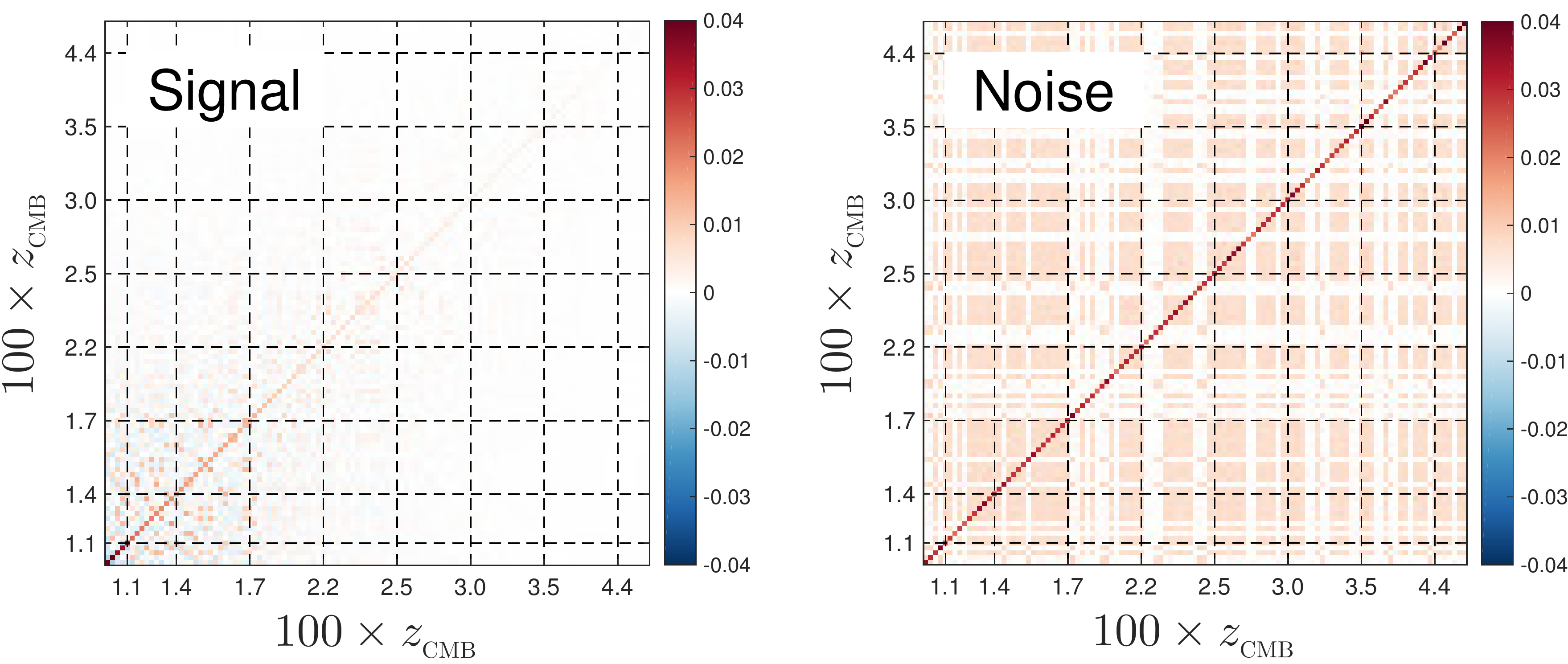}
\caption{Comparison of the signal (left panel) and noise (right panel)
  contributions to the full covariance matrix for the 111~SNe at $z < 0.05$
  from the JLA compilation.}
\label{fig:covmats}
\end{figure}

We emphasize that, since the covariance depends on the parameter $A$ that we
are interested in constraining, we need to include a term for the
$1/\sqrt{|\mathbf{C}|}$ prefactor in addition to the usual $\chi^2$
quantity. Since the covariance is a strictly increasing function of $A$,
neglecting the prefactor would lead to the clearly erroneous result that the
likelihood is a maximum for $A \rightarrow \infty$.

The likelihood in eq.~\eqref{eq:like} is the principal tool we will use for our
analyses. In this most general form, the likelihood depends on two input
quantities (four parameters, since the velocity has three components): the
normalization $A$ of the signal component of the covariance matrix and the
excess  bulk velocity $\vbulk$ not captured by the velocity
covariance. Note that, in the fiducial model, $A = 1$ and $\vbulk = 0$.

Throughout our analyses, we assume a flat $\Lambda$CDM model ($w = -1$,
$\Omega_k = 0$) with free parameters fixed to values consistent with data from
\textit{Planck} \cite{Ade:2015xua} and other probes. That is, we fix $\Omega_m
= 0.3$, physical matter density $\Omega_m h^2 = 0.14$, physical baryon density
$\Omega_b h^2 = 0.0223$, scalar spectral index $n_s = 0.965$, and amplitude of
scalar fluctuations $A_s = 2.22 \times 10^{-9}$. The corresponding derived
value of the Hubble constant is $h = 0.683$, and that of the amplitude of mass
fluctuations is $\sigma_8 = 0.79$. Within the $\Lambda$CDM model, these
parameters are determined very precisely using \textit{Planck} data alone, and
we have explicitly checked that modest changes in the cosmological model,
larger than those allowed by \textit{Planck}, have a negligible effect on our
results. We can therefore conclude that adding \textit{Planck} priors and
marginalizing over these parameters would not significantly affect our
constraints.  This is not surprising.  The background cosmology only affects
the monopole of the Supernova magnitudes, and even this dependence is weak at
very low redshifts (since we marginalize over $\mathcal{M}$).  Only a much
larger change in the parameters would affect the expected pairwise covariance,
which we do not expect to be able to measure precisely in the first place.

In figure~\ref{fig:covmats}, we compare the noise covariance $\mathbf{N}$ to
the signal covariance $\mathbf{S}$ for our fiducial cosmology. While the noise
contribution is typically larger than the signal, the signal is not
negligible, and it actually dominates for the lowest-redshift SNe. The noise,
unlike the signal, becomes effectively smaller as more SNe are used in the analysis,
making the signal important for the whole redshift range considered (see
  also the discussion in \cite{Ben-Dayan:2014swa}).

\section{Constraints on the amplitude of signal covariance}
\label{sec:A}

We first consider whether the data itself shows a preference for the presence
of the velocity (signal) covariance. Therefore, we explore the constraints on
$A$ using the likelihood in eq.~\eqref{eq:like} and fixing $\vbulk = 0$.

\begin{figure}[t]
\centering
\includegraphics[scale=0.6]{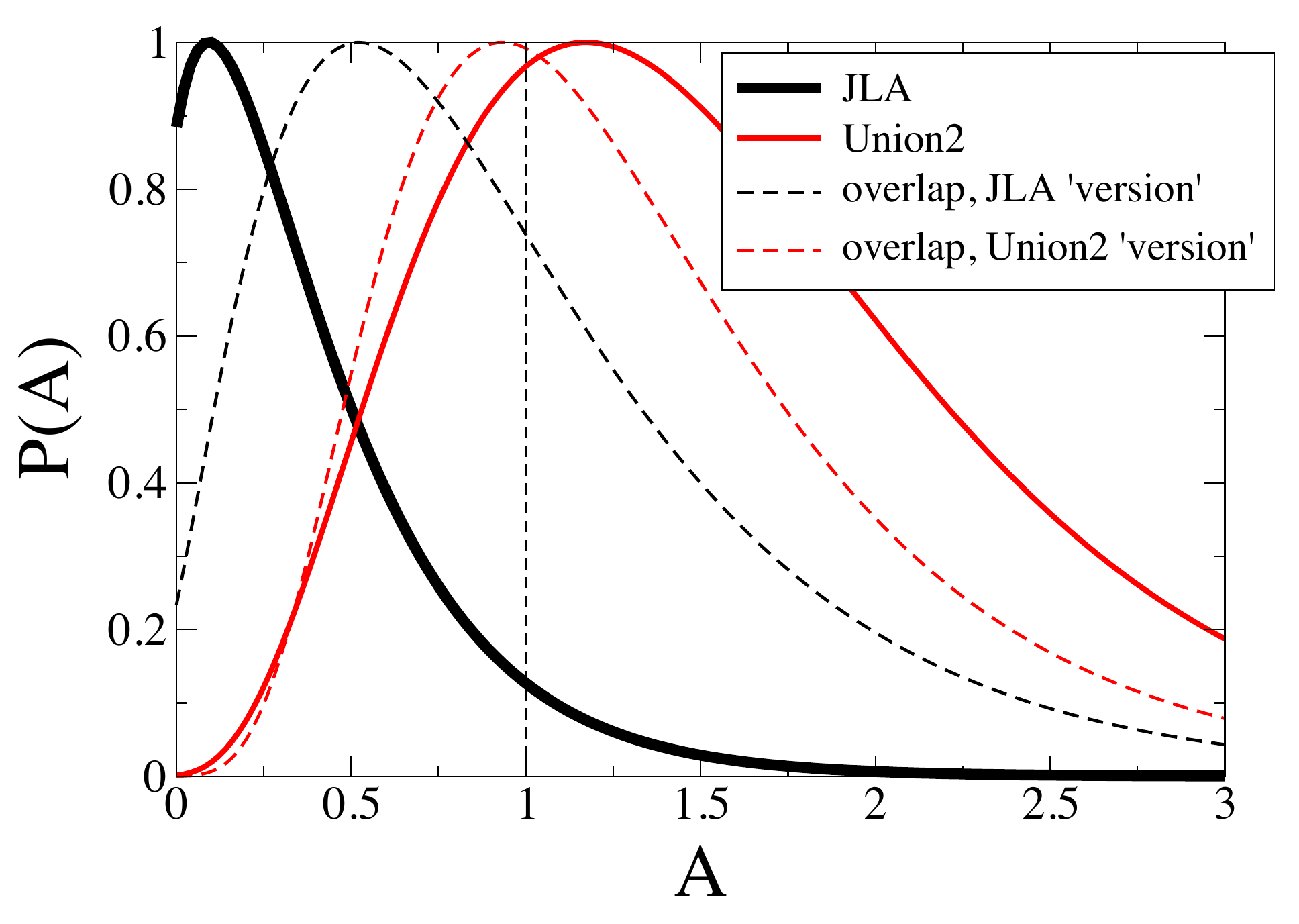}
\caption{Constraints on the parameter $A$ that quantifies the amount of
  velocity correlations ($A = 1$ is the standard $\Lambda$CDM value). The JLA
  data are consistent with $A = 1$ but do not rule out the noise-only
  hypothesis $A = 0$. JLA and Union2 give somewhat different constraints,
  though they are not statistically inconsistent. Note that differences remain
  even after restricting to the rather large subset of SNe that they have in
  common (dashed lines).}
\label{fig:like_of_A}
\end{figure}

The constraints on the parameter $A$, which determines the fraction of the
velocity covariance added to the full covariance $\mathbf{C}$, are shown in
figure~\ref{fig:like_of_A}, with the numerical results given in table~\ref{tab:summary}. We have adopted a uniform prior on $A$ such that
our Bayesian posterior is proportional to the likelihood in
eq.~\eqref{eq:like}. All data choices consistently use the available SNe out
to $z = 0.05$. This leaves 111 objects in the JLA analysis and 132 in the
Union2 analysis. We have explicitly checked that the results are insensitive
to the precise redshift cutoff; they are driven by the lowest-redshift SNe,
and $z < 0.05$ comfortably captures all of them.

The solid black curve shows JLA, the most current and rigorously calibrated
dataset. JLA does not rule out the $A = 0$ hypothesis; in fact, the likelihood
peaks near this value. Nevertheless, JLA is fully consistent with the standard
value $A = 1$, with a probability of 0.07 for $A > 1$.

The solid red curve shows the result from the Union2 dataset. While it is
noticeably different than the JLA result, the two likelihoods are mutually
consistent; in particular, $A = 1$ is a satisfactory fit to
both. Nevertheless, Union2 is different in that it strongly disfavors $A = 0$.

In order to gain additional insight into the difference between the two
datasets, we have identified SNe at $z < 0.05$ that overlap between the two
datasets, a total of 96 objects. Performing the analysis on this overlap
(dashed lines in figure~\ref{fig:like_of_A}), we see that the results are in
better agreement but still somewhat disagree, despite both analyses using the
same SN set. Part of the reason is that JLA and Union2 determine the
magnitudes differently; however, even some redshifts do not match. We find a
root-mean-square (rms) redshift difference of 1.4\% for the object-to-object
comparison of the 96 overlapping SNe, and the largest difference is 5\%. A
further exploration of precisely why the SN redshifts and magnitudes differ is
beyond the scope of this study, but we have explicitly checked that the
difference between JLA and Union2 for the overlap set is largely due to
differences in the estimated apparent magnitudes, not the
redshifts.\footnote{We have checked that the sky positions of JLA and Union2
  overlapping SNe do precisely agree. We have also found that the subset of
  these SNe that are also found in the sample of \cite{Jha:2006fm} --- 40 in
  total --- have redshifts that mostly agree very well with JLA, but show
  larger discrepancies with Union2.}

We also make connection to previous work in \cite{Gordon:2007zw}, where the
first cosmological constraints from the correlations of SN Ia peculiar
velocities have been obtained. Instead of parametrizing the covariance with a
multiplicative amplitude, they jointly constrained the cosmological parameters
$\Omega_m$ and $\sigma_8$. Using the earlier SN dataset from
\cite{Jha:2006fm}, they found constraints broadly consistent with $\Lambda$CDM
values. Using the same dataset, we roughly agree with \cite{Gordon:2007zw},
our likelihood favoring $A \simeq 1.4$ with a large uncertainty but
effectively ruling out $A = 0$. However, we note that the dataset of
\cite{Jha:2006fm} includes some SNe with extremely low redshifts (as low as $z
= 0.002$), which are not in the Hubble flow and for which the assumption of
small residuals in eq.~\eqref{eq:deltam} breaks down (see also discussion
after eq.~\eqref{eq:xi}). Objects at such extremely low redshifts are mutually
separated by distances of a few tens of Mpc; their relative velocities are
therefore expected to have important nonlinear corrections, in addition to the
linear relation eq.~\eqref{eq:deltam} breaking down, and the analysis would
have to be carefully generalized to take this into account. When we exclude
all SNe with $z < 0.01$ from the dataset of \cite{Jha:2006fm}, the likelihood
for $A$ actually looks very similar to the JLA constraints in
figure~\ref{fig:like_of_A}, favoring $A = 0$ but still statistically
consistent with $A = 1$.

To summarize, we find that JLA, the most current and rigorous dataset, does
not favor the presence of SN velocity covariance guaranteed in the
$\Lambda$CDM model, but is nonetheless consistent with it. We also find that
there is considerable variation in the SN data in terms of their constraints
on the velocity covariance, and in particular that the optimistic-looking
results found in \cite{Gordon:2007zw} were due to some very-low-redshift
  SNe that may be too nearby for accurate modeling with linear theory.

The covariance of SN flux residuals has the potential to provide additional
information about cosmological parameters and other interesting physics
\cite{Cooray:2005yp, Hannestad:2007fb, Abate:2008au,
  Bhattacharya:2010cf,Castro:2014oja}. Our results suggest that current data
do not yield interesting constraints. This will
likely change with larger, homogeneous samples with greater sky coverage, such
as those expected from the Large Synoptic Survey Telescope (LSST), currently
under construction.

\section{Constraints on excess bulk velocity}
\label{sec:vbulk}

The optimal way to search for the effect of SN
velocity correlations in the context of a fiducial cosmological model is to
test for the presence of the full signal covariance, as we did in
section~\ref{sec:A}. However, we can also use the SN magnitude residuals to
search for a dipolar distortion (for example, due to bulk motion) beyond what
is expected in $\Lambda$CDM.  This provides constraints on physics beyond the
concordance cosmology, such as a breaking of statistical isotropy or
homogeneity, or the presence of a single large long-wavelength perturbation.
The search for bulk flows is the subject of a significant body of literature.
Here we use a different approach, and we make the connection to previous
literature in the next section.

Bulk velocity is usually defined as the motion of the volume spanned by SNe Ia
and the rest frame defined by the CMB. Moving the SN redshifts to the CMB
frame, we are looking for an overall motion between the SN volume and the rest
frame. Since we aim to search for an excess bulk flow \emph{beyond} $\Lambda$CDM, we
include the velocity correlations $\mathbf{S}$ in the likelihood, fixing $A =
1$. Since $\mathbf{S}$ includes all velocity correlations within $\Lambda$CDM,
including the dipole, we expect the posterior for the bulk flow to be
consistent with zero if $\Lambda$CDM provides a good description of the SN
data. Of course, this assumes that our linear modeling of the velocity
correlations is accurate for the SN sample and also that there are no
unaccounted-for systematic errors in the SN data that
could masquerade as an excess bulk flow.

\begin{figure}[t]
\centering
\includegraphics[scale=0.6]{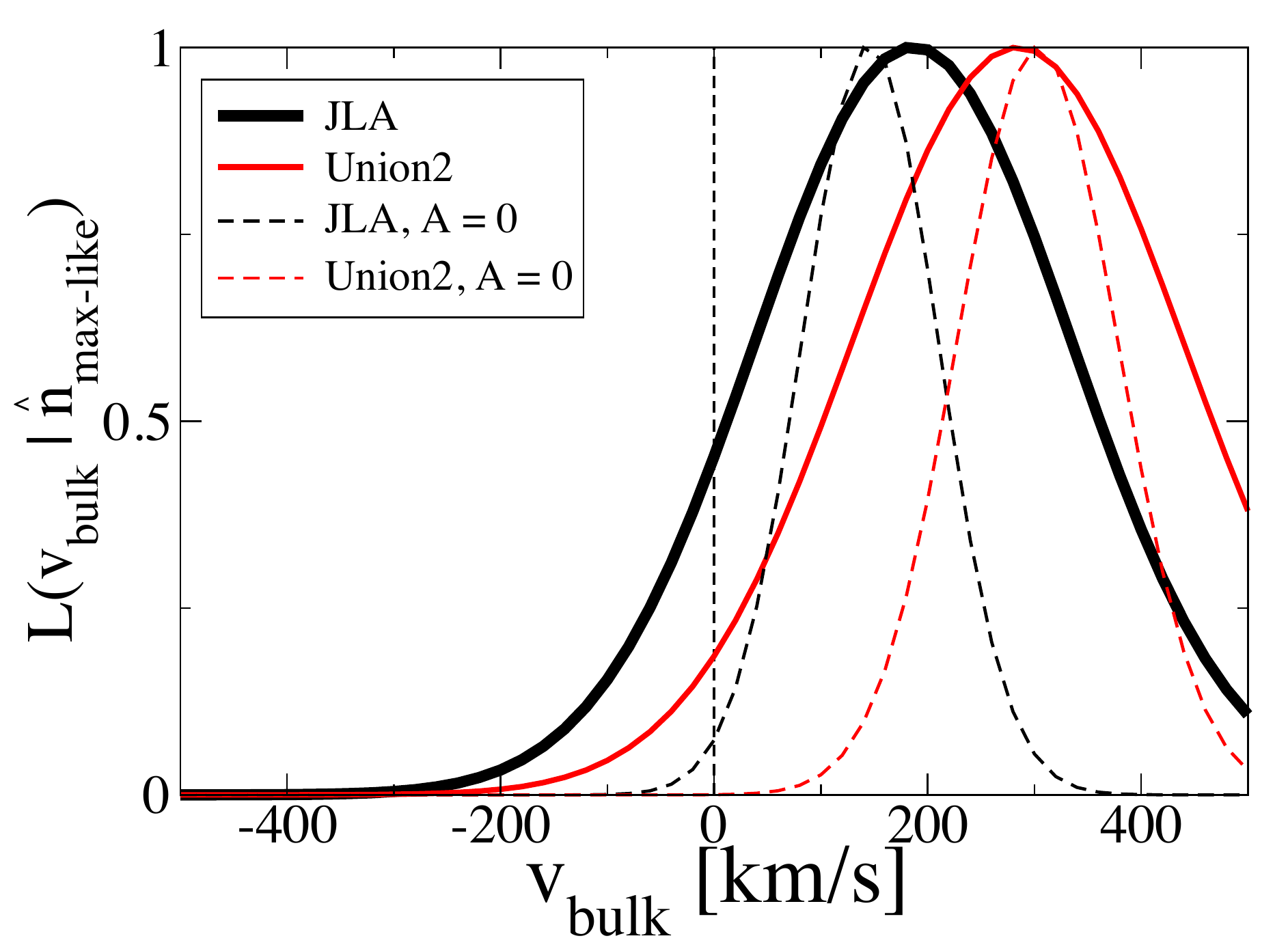}
\caption{A slice through the 3D likelihood for excess bulk velocity. The
  direction is fixed to be $\nhat_\text{max-like}$ (different for each
  dataset), while the amplitude of the dipole is varied and allowed to be
  positive or negative. Conclusions about the bulk flow would differ
  significantly if the velocity signal covariance were set to zero (dashed
  lines), as in most previous work on the subject. }
\label{fig:slice_L_of_v_best_dir}
\end{figure}

For this analysis, we therefore adopt the likelihood from eq.~\eqref{eq:like},
setting $A = 1$ but allowing the bulk velocity $\vbulk = (\vb, \theta, \phi)$
to vary in magnitude and direction. In order to get a sense of the
three-dimensional likelihood $\mathcal{L}(\vb, \theta, \phi)$, we first
consider the likelihood of the excess bulk flow amplitude in a cut through the
best-fit direction, that is, as a function of $\vb$ with $\theta$ and $\phi$
set to their maximum-likelihood values. This is shown in
figure~\ref{fig:slice_L_of_v_best_dir}, and note that we continue the scan
past zero velocity in the direction opposite that of the best fit by letting
the amplitude of the bulk flow take negative values. Because the likelihood is
Gaussian, and because $\vbulk$ enters the observable magnitude linearly (see
eq.~\eqref{eq:dm}), the likelihood of the bulk velocity is also guaranteed to
be Gaussian. Therefore, the likelihood ratio between the best-fit $(\vb,
\theta, \phi)$ and $\vb = 0$ immediately translates into the confidence at
which zero excess bulk velocity is ruled out, assuming uniform priors on each
component of the vector $\vbulk$.

Figure~\ref{fig:slice_L_of_v_best_dir} shows that, once the velocity
covariance is properly taken into account, SN data do not favor any bulk
velocity beyond the amount expected in $\Lambda$CDM. For example, the JLA
likelihood peaks at $(\vb, l, b) = (187~\text{km/s}, 323^\circ, 25^\circ)$,
but this likelihood is larger than that for $\vb = 0$ only by
$-2\Delta\ln\mathcal{L} = -1.6$, far too low even for 68.3\% (1$\sigma$)
evidence, which in three dimensions would be $-2\Delta\ln\mathcal{L} \simeq
-3.5$.

We would now like to explicitly calculate the posterior distribution of the
amplitude of an excess bulk velocity. Assuming uniform priors on each
component of $\vbulk$ would produce an additional $\vb^2$ factor in the
posterior, driving it to zero for the $\vb = 0$ case so that $\vbulk = 0$
  is automatically ruled out. An alternative that allows us to test the
$\vbulk = 0$ assumption, implicitly (or explicitly \cite{Ma:2010ps}) adopted
by most previous work, is to choose the prior $\text{Pr}(\vb) \propto 1/\vb^2$
or, alternatively, to consider the angle-averaged likelihood
\begin{equation}
P^\text{angle-averaged}(\vbulk)
\propto \int \mathcal{L}(\vbulk) \, d\cos\theta \, d\phi \, .
\label{eq:like_without_vsq}
\end{equation}

\begin{figure}[t]
\centering
\includegraphics[scale=0.6]{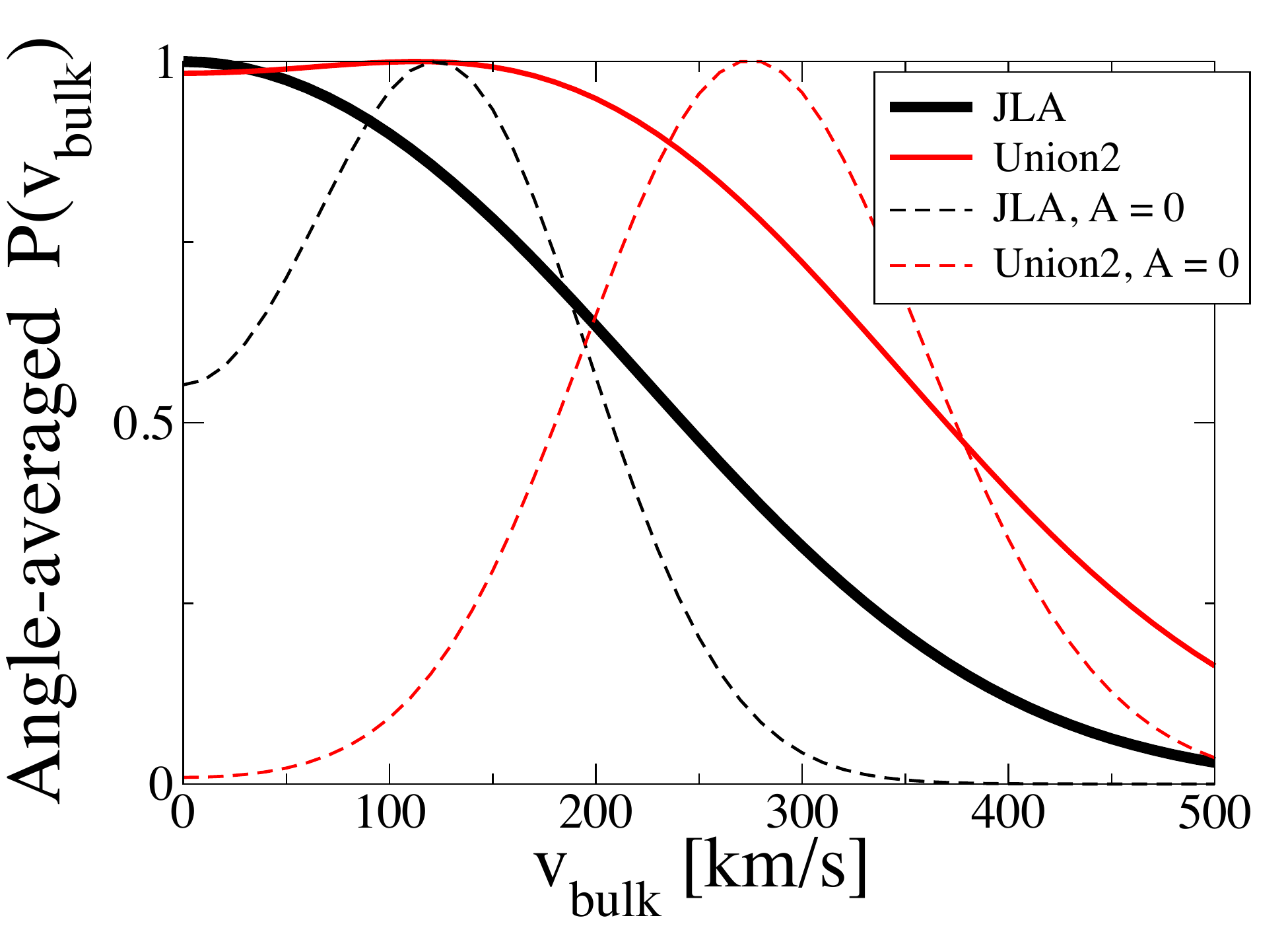}
\caption{Angle-averaged posterior on the amplitude of bulk velocity in excess
  of the correlations captured by the full $\Lambda$CDM velocity
  covariance. Both JLA and Union2 data are consistent with zero velocity, with
  relatively large error. The conclusion would again be very different if the
  velocity covariance were artificially set to zero (dashed lines).}
\label{fig:like_of_v}
\end{figure}

We plot this likelihood in figure~\ref{fig:like_of_v}. It is immediately
apparent that both JLA and Union2 data show no preference for excess bulk
velocity, though there is a large uncertainty.  Performing the same analysis
but setting the velocity correlations to zero (so that $\mathbf{C}
=\mathbf{N}$ with $\mathbf{S} = 0$), the results are drastically different,
favoring bulk flows of several hundred km/s and, in the case of Union2, firmly
ruling out the $\vb = 0$ case. This is in agreement with the conclusion found
in previous work \cite{Watkins:2008hf, Colin:2010ds, Dai:2011xm,
  Turnbull:2011ty, Feindt:2013pma, Mathews:2014fma}. In the dashed curves
shown in figure~\ref{fig:like_of_v}, we have not included the 150~km/s
(300~km/s) scatter that is added in quadrature to the diagonal of the noise
covariance $\mathbf{N}$ for JLA (Union2) data in some analyses. This clearly
does not capture the significant full covariance of SN residuals due to
large-scale structure. However, when adding back this contribution, we find
nearly perfect agreement with the results of \cite{Dai:2011xm}.

Given the importance of this issue, we stress again that the dashed lines in
figures~\ref{fig:slice_L_of_v_best_dir} and \ref{fig:like_of_v} \emph{do not
  show the proper likelihood of any peculiar velocity}, $\Lambda$CDM or
otherwise, since a guaranteed component of the covariance of the data has been
neglected (nevertheless, one can still use these incorrect likelihoods to
construct an estimator for bulk flow, which we will turn to in the next
section).  Since the velocity covariance $\mathbf{S}$ gives a guaranteed
source of correlations in $\Lambda$CDM, we argue that it should always be
included in likelihood analyses of SN magnitude residuals.  Neglecting this
covariance will lead to suboptimal estimators and, in general, biased results.

\section{Relation to previous bulk flow measurements}
\label{sec:vbulk_noise}

A significant body of earlier work on SN velocities neglected the velocity
covariance (and the lensing covariance, which is important at higher
redshifts). These analyses often found evidence for non-zero bulk flow, and we
confirm these findings with our dashed curves in
figures~\ref{fig:slice_L_of_v_best_dir} and \ref{fig:like_of_v}. This bulk
velocity with $A = 0$ is difficult to interpret, since it was obtained in an
analysis with a guaranteed contribution to the covariance of the observables
artificially set to zero.

Nevertheless, one can derive a theoretical expectation for what one should
expect for the bulk velocity derived in this way; we call it
$\vbulk^\text{noise-only}$, and in some previous work it corresponds to what
the authors simply call ``bulk velocity.''  Note that this is not the
\emph{excess} bulk velocity over the $\Lambda$CDM expectation considered in
the previous sections.

We will adopt the likelihood from
eq.~\eqref{eq:like} once more, but we define a new vector $\mathbf{x}$ via
$\mathbf{\Delta m}^\text{bulk}(\vbulk) \equiv \vb \, \mathbf{x}$, where

\begin{equation}
x_i \equiv -\left(\frac{5}{\ln 10} \right) \, \frac{(1 + z_i)^2}{H(z_i) d_L(z_i)} \ \nhat_i \cdot \nhat_\text{bulk} \, .
\end{equation}

To estimate $\vb^\text{noise-only}$, we set the signal covariance to zero,
assume a fixed direction $\nhat_\text{bulk}$, and find the maximum of the
likelihood; that is
\begin{equation}
\widehat{v}_\text{bulk}^\text{noise-only} \longleftrightarrow \, \underset{\vb}{\text{max}} \, \left[\mathcal{L}(A = 0, \vbulk) \right] \, .
\end{equation}
Maximizing with respect to $\vb$, one finds
\begin{equation}
\widehat{v}_\text{bulk}^\text{noise-only} = \frac{\mathbf{y}^\intercal \, \mathbf{\delta m}}{\mathbf{y}^\intercal \, \mathbf{x}} \, ,
\end{equation}
where we have defined for convenience a new vector $\mathbf{y}$ as $\mathbf{y}
\equiv \mathbf{N}^{-1} \mathbf{x}$, and where $\mathbf{\delta m}$ is the
vector of SN magnitude residuals (eq.~\eqref{eq:dm} without the $\Delta
m^\text{bulk}$ term). Assuming $\Lambda$CDM, and still keeping the direction
$\nhat_\text{bulk}$ fixed, the expected (mean) velocity is of course zero
since $\langle \delta m_i \rangle = 0$.

We are more interested in the variance of this quantity, which can be computed
directly and is equal to
\begin{equation}
\left\langle \left(\widehat{v}_\text{bulk}^\text{noise-only} \right)^2
\right\rangle =
\frac{\mathbf{y}^\intercal (\mathbf{S} + \mathbf{N}) \, \mathbf{y}}{(\mathbf{y}^\intercal \, \mathbf{x})^2} \, ,
\end{equation}
since the true SN magnitude covariance for $\Lambda$CDM is the sum of both
signal and noise: $\langle (\delta m_i) (\delta m_j) \rangle = S_{ij} +
N_{ij}$. The square root of this quantity is the desired theoretical
expectation for the rms bulk velocity in $\Lambda$CDM when one ignores the signal covariance
matrix.

Using the JLA SNe up to redshift $z = 0.05$, the $\Lambda$CDM expectation for
the rms velocity varies from about 150--170~km/s as a function of
$\nhat_\text{bulk}$, with a sky-averaged value of 162~km/s. Assuming only
noise in the data, $\langle (\delta m_i)(\delta m_j) \rangle = N_{ij}$, the
result is 71~km/s. The Union2 data give similar results.

We have therefore found that the predicted rms value of
$\vb^\text{noise-only}$, assuming $\Lambda$CDM and SN data up to $z = 0.05$,
is $\sim$160~km/s, and that nearly half of this value would be generated by
statistical scatter in SN magnitudes in the absence of any peculiar
velocities in the universe (such a contribution is sometimes referred to as
the ``noise bias'').  The peak of the JLA likelihood (black dashed line in
figure~\ref{fig:like_of_v}) is in agreement with the $\Lambda$CDM expectation;
Union2 gives a somewhat larger result. Again, however, we caution that the
analysis in this section is suboptimal, given that we do condense the data
into a weighted dipole estimator $\widehat{v}_\text{bulk}^\text{noise-only}$
rather than using the full covariance.  Moreover, this estimator is
significantly affected by noise bias which needs to be subtracted.  Given the
uncertainties in the noise covariance, the subtraction of noise bias will lead
to additional systematic uncertainties in the actual peculiar velocity
measurement.

We have not attempted to repeat the analyses of some past work that studied
the velocity field of low-redshift SNe
\cite{Haugboelle:2006uc,Turnbull:2011ty,Weyant:2011hs,Ma:2012tt,Feindt:2013pma,Johnson:2014kaa,Ma:2013oja,
  Mathews:2014fma} or the anisotropy of the universe from the distribution of
nearby SN distances \cite{Schwarz:2007wf,
  Kalus:2012zu, Yang:2013gea, Appleby:2014kea, Lin:2015rza, Javanmardi:2015sfa},
since these studies adopted a wide variety of approaches and, in some cases,
complicated statistical procedures whose results are calibrated on
simulations. We emphasize, however, that the velocity covariance due to
large-scale structure should be included in any such analyses in order to
obtain unbiased results and draw reliable statistical conclusions about the
velocity field of the nearby universe.

\renewcommand{\arraystretch}{1.5}
\begin{table*}[!t]
\centering
\begin{tabular}{|c|cccc|cc|c|}
\hline
& \multicolumn{4}{c|}{$P(A)$ (fig~\ref{fig:like_of_A})} & \multicolumn{2}{c|}{$\vb^\text{max-like}$ (fig~\ref{fig:slice_L_of_v_best_dir})} & \multicolumn{1}{c|}{$\vb^\text{angle-avg}$ (fig~\ref{fig:like_of_v})} \\
Survey & ML & 95\% C.L. & $\left.\Delta\chi^2\right|_{A = 0}$ & $\left.\Delta\chi^2\right|_{A = 1}$ & ML & 95\% C.L. & 95\% C.L. \\
\hline
JLA & 0.19 & $(0, 1.15)$ & 0.24 & 4.13 & 187 & $(-108, 485)$ & $(0, 376)$ \\
Union2 & 1.19 & $(0.19, 3.27)$ & 13.2 & 0.07 & 265 & $(-37, 568)$ & $(0, 456)$ \\
\hline
\end{tabular}
\caption{Summary of numerical results. For both JLA and Union2, we show the
  best-fit, maximum-likelihood (ML) values and 95\% confidence intervals for
  $A$. We also show the quantity $\Delta\chi^2$ (that is, $-2\Delta\ln L$)
  between the best-fit value and special values $A = 0$ (no velocity signal)
  and $A = 1$ ($\Lambda$CDM velocity signal). We also show ML and 95\%
  intervals for bulk velocity in the best-fit direction and 95\% intervals for
  angle-averaged bulk velocity (here we do not report ML values, which are
  near zero). All velocities are in units of km/s.}
\label{tab:summary}
\end{table*}

\section{Conclusions}
\label{sec:concl}

In this paper we have revisited the constraints on bulk velocity --- the
relative motion of the volume populated by nearby SNe Ia and the rest frame
defined by the CMB. Our emphasis was on a precise and clear procedure for
selecting the data, performing the analysis, and modeling the theoretical
expectation. We concentrated on SNe Ia as tracers of cosmic structure and
studied two separate (but overlapping) datasets. Our methodology applies
equally well to galaxies and other tracers of large-scale structure.

We argued, and demonstrated with explicit calculations, that inclusion of the
``signal'' covariance matrix that captures the peculiar velocity correlations
between SNe is crucial. The velocities provide a guaranteed source of
covariance between SNe; while the velocity contribution is subdominant
compared to the noise except at the lowest redshifts (see
figure~\ref{fig:covmats}), it does not become smaller as more SNe are included
in the analysis.  Neglecting the velocity covariance, as done by a significant
body of earlier work on SN velocities and tests of statistical isotropy, leads
to results that are both biased and difficult to interpret.

Our approach was based on a likelihood that includes both the signal and noise
covariance and four free parameters: the normalization $A$, specifying the
fraction of the signal added to the covariance, and three components of an
excess bulk velocity $\vbulk$ beyond that which is encoded in the signal
covariance. For the fiducial $\Lambda$CDM model, $A = 1$ and $\vbulk = 0$.

We first investigated whether the standard $\Lambda$CDM velocity covariance
($A = 1$) is preferred over the case in which the covariance is ignored ($A =
0$); that is, we constrained $\mathcal{L}(A, \vbulk = 0)$. We found that the
JLA dataset, while consistent with $A = 1$, cannot rule out $A = 0$, and in
fact its likelihood peaks near zero (figure~\ref{fig:like_of_A}). Therefore,
we did not find convincing evidence in the data for the correlations expected
from velocities. Although we expect things to change with future data, when
precise measurements of a quantity like $A$ will effectively constrain
cosmological parameters such as $\Omega_m$ and $\sigma_8$, our results
indicate that current data are not close to providing such useful constraints.

We then pursued a different approach; we assumed a standard $\Lambda$CDM
velocity covariance (the $A = 1$ case) and tested for excess bulk velocity
$\vbulk$ beyond that already captured by the covariance --- that is, we
adopted the likelihood $\mathcal{L}(A = 1, \vbulk)$ in the analysis. We found
that current SN data provide no evidence for a departure from the null
hypothesis $\vbulk = 0$ (figures~\ref{fig:slice_L_of_v_best_dir} and
\ref{fig:like_of_v}). This result is in sharp contrast to the inference one
would have made by ignoring the velocity covariance (that is, setting $A = 0$
in the same analysis), as some previous analyses in the literature have done.

To better understand this latter case, we separately studied inferred
constraints on a ``non-excess'' bulk velocity where the velocity covariance has been
ignored --- that is, the $\mathcal{L}(A = 0, \vbulk^\text{noise-only})$
case. Note that this bulk velocity is more difficult to interpret since it was
obtained by ignoring a guaranteed source of correlations in the data. We
showed that the rms of the estimated $\vb^\text{noise-only}$, assuming
$\Lambda$CDM and SN data up to $z = 0.05$, is expected to be $\sim$160~km/s,
and that nearly half of this
value would be generated by a contribution purely from
intrinsic and observational scatter in the SN magnitudes. Therefore, there are really two problems
with this approach: not only is the constrained quantity difficult to
interpret, but it is also guaranteed to be nonzero even without any peculiar
velocities in the universe, which is clearly not optimal for cosmological
interpretations.

The mapping of velocities in the universe using nearby tracers of large-scale
structure has had a remarkably long and productive history. With upcoming
large-field, fast-scanning surveys, it is likely that data will become of
sufficiently high quality to enable peculiar velocities to progress to the
next level and become competitive cosmological probes. Of course, data analysis and
theoretical modeling will have to progress as well. In this paper we have
demonstrated that, even for current data, clearly defining the
quantities to be constrained and carefully accounting for the 
guaranteed correlations between objects due to large-scale structure
are two factors of key importance.

\acknowledgments We thank Alex Conley, De-Chang Dai, Vera
Glu\v{s}\v{c}evi\'{c}, Chris Gordon, Will Kinney, Yin-Zhe Ma, and An\v{z}e
Slosar for useful discussions and Greg Aldering, Mike Hudson, and Alex Kim for
comments on an earlier draft of this paper. We also thank the anonymous
  referee for useful comments. DH and DLS are supported by NSF
under contract AST-0807564 and DOE under contract DE-FG02-95ER40899. DH has
also been supported by the DFG Cluster of Excellence ``Origin and Structure of
the Universe'' (\url{http://www.universe-cluster.de/}).

\bibliography{snvel}

\end{document}